\documentclass[prl,twocolumn, aps,preprintnumbers,showpacs, superscriptaddress]{revtex4}

\usepackage{latexsym}
\usepackage{amsmath}
\usepackage{amssymb}
\usepackage{revsymb}
\usepackage{graphicx}
\usepackage{amsthm}

\usepackage{tikz}
\usetikzlibrary{matrix,shapes.geometric,arrows,positioning,chains,scopes,calc,through}

\newcommand{\abs}[1]{\left|#1\right|}
\newcommand{\be}{\begin{equation}}
\newcommand{\ee}{\end{equation}}
\newcommand{\ba}{\begin{array}}
\newcommand{\ea}{\end{array}}
\newcommand{\bea}{\begin{eqnarray}}
\newcommand{\eea}{\end{eqnarray}}
\newcommand{\Tr}[1]{\mbox{Tr}\left[#1\right]}

\newcommand{\rank}{\mathop{\mathrm{Rank}}\nolimits}

\newtheorem{prop}{Proposition}
\newtheorem{theorem}{Theorem}

\begin{document}

\title{Qubit State Discrimination}

\author{Matthieu E. \surname{Deconinck}}
\email[]{deconinc@telecom-paristech.fr} \affiliation{TELECOM
ParisTech, 37/39 rue Dareau, 75014 Paris, France}
\author{Barbara M. \surname{Terhal}}
\email[]{bterhal@gmail.com} \affiliation{IBM Watson Research Center,
P.O. Box 218, Yorktown Heights, NY, USA}
\date{\today}

\begin{abstract}
We show how one can solve the problem of discriminating between 
qubit states. We use the quantum state discrimination duality
theorem and the Bloch sphere representation of qubits which allows
for an easy geometric and analytical representation of the optimal guessing
strategies.
\end{abstract}

\pacs{03.67.Ac, 89.70.Eg, 31.15.am}

\maketitle

A fundamental task in quantum information theory is that of quantum
state discrimination (QSD). It is the problem of distinguishing between
the possible states of a physical system. Generally, state
discrimination is not a task that can be achieved perfectly; even
classically, the bias of a die cannot be determined with certainty by throwing it once. Moreover, a characteristic feature of quantum mechanics is the impossibility to distinguish non-orthogonal pure states with certainty. Thus, one problem of quantum state discrimination is to
determine which measurement strategies maximize the probability $P_{\rm guess}$ of correctly 
guessing the state. Such measurement strategies will be called optimal strategies.
Other figures of merit have been used to measure the success in identifying a quantum state, for example one can consider the task of unambiguously distinguishing among quantum states.
However, it is the maximum guessing probability $P_{\rm guess}$, and the so-called min-entropy equal to $-\log P_{\rm guess}$, which has appeared as the essential quantity 
which measures uncertainty about classical information in quantum cryptography, see e.~g.~\cite{minentropy}.

Since the problem of quantum state discrimination is so
fundamental to quantum physics, many results have been
obtained over the last 40 years. The most important of these is the
formulation of the optimal guessing strategies given a
set of probable quantum states, as a semi-definite program \cite{Holevo-Yuen, EMV:semidef} (see also \cite{BC:sdp}). The semi-definite program formulation shows that the quantum state discrimination problem is computationally tractable. The formulation also dashes any hope for an analytical
solution to the quantum state discrimination problem in general. The
special cases which have allowed for particular or closed-form answers so far
include the well-known case of two states \cite{Helstrom2} and the
scenario where the states themselves can be used to compose the
POVM elements of an optimal measurement \cite{ykl:self}. Another important subset of problems constitute those families of quantum states for whom Belavkin's square-root measurement (which was later reinvented as the pretty good measurement \cite{lsm0}) or Belavkin's weighted square-root measurement \cite{belav_radio,belav_optimal} is optimal. We refer to \cite{holevo, sasaki, ban_etc, EMV:semidef, mochon} and \cite{tyson_error} for various results on the use of the (weighted) square-root measurement.


The last set of solution strategies pertains to the particular
problem of distinguishing between qubit states. Ref. \cite{qubits}
analyzes the task of optimally distinguishing between pure (linearly-dependent)
qubit states and finds partial results using the Bloch sphere
representation. Other partial results on distinguishing pure or 
equiprobable qubit states were obtained in \cite{mirror, Hunter, Hunter2}.
We refer to \cite{SoA} for a current overview of the quantum state discrimination problem.



What is lacking in the qubit state discrimination problem is a systematic investigation of how the necessary and sufficient conditions of the quantum state discrimination duality theorem can be employed to determine the optimal strategies for general sets of qubit states.  In this Letter we provide such simple analysis. For qubits the semi-definite program of QSD reduces to a program for finding the minimal enclosing ball of a set of balls in $\mathbf{R}^3$. For this well-studied problem, there exists an algorithm whose running time is {\em linear} in the number of states to be distinguished. Such linear-time algorithm should be contrasted with the polynomial but not necessarily very efficient complexity of a semi-definite program. 
In addition, it is not hard to show that for qubits there is an optimal strategy which involves distinguishing at most a subset of four states. We will describe an entirely {\em analytical} procedure for determining the optimal strategy for discriminating between four states.

We start by recalling the quantum state discrimination duality
theorem. Let a state discrimination problem
$\left(\mathcal{X},\mathcal{H},\mathcal{R}\right)$ be defined by a
finite alphabet $\mathcal{X}$ with $|\mathcal{X}|=m$, an
$n$-dimensional Hilbert space $\mathcal{H}$ and a set of positive
semi-definite operators on this Hilbert space,
$\mathcal{R}=\left(\rho_x\geq 0\right)_{x\in \mathcal{X}}$ such
that $\Tr{\sum_{x \in \mathcal{X}}\rho_x}=1$. For such a problem, a
guessing strategy is defined by a measurement
$\mathcal{E}=\left(E^x \geq 0\right)_{x\in\mathcal{X}}$ with $\sum_{x
\in \mathcal{X}} E^x=I_{\mathcal{H}}$. The probability of correctly
guessing $x$ using the POVM $\mathcal{E}$ is given by
$p(\mathcal{E})=\sum_{x\in\mathcal{X}}\Tr{E^x\rho_x}$.
Alternatively, we can associate the operator
$\sigma(\mathcal{E})=\sum_x E^x \rho_x$ with a strategy
$\mathcal{E}$ such that the guessing probability equals
$p(\mathcal{E})=\Tr{\sigma(\mathcal{E})}$. Hence, the problem of
determining the optimal guessing strategies is to find an operator
$\sigma$ and its associated measurement strategies which has maximal
trace $P_{\rm guess}$.

\begin{theorem}[QSD Duality Theorem \cite{Holevo-Yuen, EMV:semidef, BC:sdp}]
Let $\left(\mathcal{X},\mathcal{H},\mathcal{R}=\left(\rho_x\right)_{x\in
\mathcal{X}}\right)$ be a discrimination problem. Then all optimal measurement strategies 
$\mathcal{E}$ define the same operator $\sigma(\mathcal{E})$ and
this operator is called the \textbf{Lagrange operator} of the problem and denoted as
$\sigma_{\mathcal{R}}$. It has the following properties:
\begin{align}
&\sigma_{\mathcal{R}}=\sigma_{\mathcal{R}}^{\dagger}, \\
&\forall x\in \mathcal{X}:\,\sigma_{\mathcal{R}}\geq\rho_x \label{order},\\
& \forall x\in\mathcal{X},\;E^x\sigma_{\mathcal{R}}=E^x\rho_x\, \Leftrightarrow\, \mathcal{E}=(E^x)_{x\in\mathcal{X}} \mbox{ is optimal.}\label{useful}
\end{align}
Alternatively, the Lagrange operator $\sigma_{\mathcal{R}}$ can be found as the solution of the following semi-definite program
\bea
\label{qsdd}
& \min \Tr{\sigma}, \nonumber & \\
& \mbox{such that }\forall x \in \mathcal{X},\; \sigma \geq \rho_x, &  \\
& \mbox{and } \sigma=\sigma^{\dagger}. &. \nonumber
\eea
Lastly, any guessing strategy $\mathcal{E}$ with operator $\sigma(\mathcal{E})=\sum_x E^x \rho_x$ such that
\be
\forall x\in \mathcal{X},\;\frac{\sigma+\sigma^{\dagger}}{2}\geq \rho_x,
\label{optimalstrat}
\ee
is an optimal strategy.
\end{theorem}

We will first consider the semi-definite program in the qubit setting. Every Hermitian operator on qubits can be written as $\rho=\frac{1}{2}(pI_{\mathcal{H}}+\vec{r}\cdot\vec{\sigma})$ where
$(p,\vec{r}) \in \mathbf{R}^4$ with $\vec{r}=\Tr{\rho\vec{\sigma}}$ and $p=\Tr{\rho}$.
Here $\vec{\sigma}=(\sigma_x, \sigma_y,\sigma_z)$
are the Pauli matrices. A special feature of $2 \times 2$ Hermitian matrices is that the condition of positive semi-definiteness can be expressed as a quadratic constraint, i.e. $p \geq 0$ and $\abs{\vec{r}}^2 \leq p^2$.
Hence, in this representation the Lagrange operator for distinguishing the set
$\left(\rho_x\right)_{x\in \mathcal{X}} \equiv \left((p_x,\vec{r}_x)\right)_{x \in \mathcal{X}}$ 
 is given by the solution $\sigma \equiv (t,\vec{s}) \in \mathbf{R}^4$ of the following program:
\bea
& \min t, \nonumber & \\
& \mbox{such that }\forall x \in \mathcal{X},\; t \geq p_x, &  \nonumber \\
& \mbox{ and } \forall x \in \mathcal{X},\;(t-p_x)^2 \geq |\vec{s}-\vec{r}_x|^2. &
\label{sigmaSDP}
\eea
This is an example of a conic quadratic program for vectors in $\mathbf{R}^4$ (conic quadratic programs form a subclass of general semi-definite programs and include linear programs \cite{BTN:cp}). The constraints in such program are specified by quadratic cones. With each vector $(p_x, \vec{r}_x)$ we can associate a ``light" cone ${\cal C}_x$ which consists of all vectors $(t,\vec{s})$ such that $(t-p_x)^2 \geq |\vec{s}-\vec{r}_x|^2$. This cone
consists of a future light cone ${\cal C}_x^+$ for which $t \geq p_x$ and past light cone ${\cal C}_x^-$ for which $t < p_x$. The Lagrange operator of a set of states $((p_x, \vec{r}_x))_{x\in\mathcal{X}}$ precisely corresponds to the point $(t,\vec{s})$ which lies in the intersection of the future light cones $\cap_x {\cal C}_x^+$ for which the time-coordinate $t$ is minimal. Equivalently, the Lagrange operator is the state $\sigma$ with minimal trace such that all quantum states to be discriminated lie in the past light cone ${\cal C}_{\sigma}^-$ of this operator. 

We wish to express the four-dimensional vectors as balls in $\mathbf{R}^3$ whose radius depends on the ``time"-coordinate which represents the trace of the density matrix. The mapping will depend on the discrimination problem, i.~e.~let $p=\max_{x \in \mathcal{X}} p_x > 0$. We represent each state $(p_x, \vec{r}_x)$, by a ball $\mathcal{B}_x$ centered at $\vec{r}_x$ with radius $|p-p_x|$. For example, a set of equiprobable states ${\cal R}$ will be mapped onto a set of points in $\bf{R}^3$.
Note that each ball $\mathcal{B}_x$ is the intersection of the (future) light cone ${\cal C}_x$ and a three-dimensional hyperplane $H_p$ with time-coordinate equal to $p$. Similarly, the ball corresponding to the Lagrange operator $\sigma$ is obtained by intersecting the past light cone of $\sigma \equiv (t,\vec{s})$ with the hyperplane $H_p$. We can then characterize the Lagrange operator as follows.

\begin{prop}
Given is a set of balls $(\mathcal{B}_x)_{x\in \mathcal{X}}$ in $\mathbf{R}^3$ corresponding to a qubit state discrimination
problem $\rho_x \equiv (p_x, \vec{r}_x)$ and let $p=\max_{x \in \mathcal{X}} p_x$. The Lagrange operator
$\sigma=\frac{1}{2}(tI_{\mathcal{H}}+\vec{s}\cdot \vec{\sigma})$
 of this discrimination problem corresponds
to the ball $\mathcal{B}_{\sigma}$ of minimum radius $t-p$ and center $S$ at $\vec{s}$ which has non-empty intersection with each $\mathcal{B}_x$, i.e. $\forall x \in \mathcal{X},\;\mathcal{B}_{\sigma} \cap \mathcal{B}_x \neq \emptyset$. This ball with minimum radius and non-empty intersection will be called the ``interior ball" of the set $(\mathcal{B}_x)_{x \in \mathcal{X}}$.
\label{descrip}
\end{prop}

The proposition follows readily: since the past light cone of the Lagrange operator $\sigma$ should contain all points $\rho_x$, it should have a non-empty intersection with each ball $\mathcal{B}_x$. The radius of the ball ${\cal B}_{\sigma}$ obtained by intersecting the cone $\mathcal{C}_{\sigma}^-$ with $\Tr{\sigma}=t$ with the hyperplane $H_p$ is $t-p$. The Lagrange operator and thus the interior ball of $(\mathcal{B}_x)_{x \in \mathcal{X}}$ obtained in this way is always unique. Note that for a general set of three-dimensional balls $\left(\mathcal{B}_x\right)_{x\in\mathcal{X}}$, the interior ball is unique if and only if the intersection $\cap_x \mathcal{B}_x$ contains at most one point.

Before we discuss how to explicitly obtain the Lagrange operator and the optimal strategies in the representation of Proposition \ref{descrip}, let us present an observation concerning the number of states $|\mathcal{X}|=m$ to be distinguished in a Hilbert space of dimension $n$. If $m$ is large, say $m \gg n$, then it may seem hard to determine the optimal strategy. In addition, an optimal strategy with $m$ outcomes may be hard to implement practically. Fortunately, it is not hard to show that there always exists an optimal strategy which has at most $n^2$ outcomes (see also \cite{Hunter}). Davies \cite{davies} has shown that any POVM measurement $\mathcal{E}$ on states in a $n$-dimensional Hilbert space can be written as a convex combination of POVM measurements each of which has at most $n^2$ outcomes (this can be viewed as a consequence of Carath\'{e}odory's theorem applied to the $n^2-1$-dimensional real vector space of trace-1 Hermitian matrices).

Since the guessing probability is a convex (linear) function of the guessing strategy, it follows that an optimal guessing strategy can be written as a convex combination of optimal guessing strategies each of which has at most $n^2$ outcomes.
 
Clearly, the Lagrange operator for this optimal strategy $\sigma=\sum_{x \in \mathcal{X}'} E^x \rho_x$ where $\mathcal{X}' \subseteq \mathcal{X}$ and $|\mathcal{X}'|\leq n^2$ obeys Eq.~(\ref{optimalstrat}) for the ensemble $(\rho_x)_{x \in \mathcal{X}'}$. Therefore this strategy is optimal for distinguishing states in the subset $\mathcal{R}'=(\rho_x)_{x\in \mathcal{X}'}$. Thus $\sigma_{\mathcal{R}'}=\sigma_{\mathcal{R}}$ and all the optimal measurement strategies for $\mathcal{R}'$ are optimal for $\mathcal{R}$. 


It follows that in order to find the Lagrange operator for the whole set of states $\mathcal{R}$, we may consider the Lagrange operators for subsets $\mathcal{R}'$ of size at most $n^2$. All Lagrange operators $\sigma_{\mathcal{R}'}$ for the subproblems $\mathcal{R}'$ have the property that $\Tr{\sigma_{\mathcal{R}'}} \leq \Tr{\sigma_{\mathcal{R}}}$ and thus the Lagrange operator $\sigma_{\mathcal{R}'}$ {\em with the largest trace} is the Lagrange operator for the whole problem. 


For qubits, these arguments show that one can obtain the Lagrange operator and the optimal measurement strategies by considering the problem of distinguishing between at most four states. Hence, if the number of states $m$ in $\mathcal{R}$ is larger than four, one may solve the state distinguishability problem by considering all ${m \choose 4}$ subsets $\mathcal{R}_4^i$ of four states and compute the Lagrange operator for each subset. Among these subsets, one chooses the subset whose Lagrange operator has maximal trace. This Lagrange operator is the Lagrange operator of the whole problem and the optimal strategy which discriminates the states in this subset is optimal for the whole problem. Below, we will present an analytical procedure for solving the qubit discrimination problem for four states. 

Note that the procedure of considering all possible subsets of four elements is not very efficient as a function of $m$. However, one can show that the particular conic program in Eq.~(\ref{sigmaSDP}) can be solved by an algorithm which runs in a time which is linear in the number of states to be distinguished. We note that Eq.~(\ref{sigmaSDP}) precisely specifies the (unscaled) Lagrange ball centered at $\vec{s} \in \mathbf{R}^3$ with radius $t$ as the ball of minimum radius which {\em includes} all (unscaled) balls $\mathcal{B}_x$ centered at $\vec{r_x}$ with radius $p_x$ \footnote{Note that this representation of the Lagrange operator is different (but equivalent in expression) from the interior ball representation of Proposition 1.}. For this particular problem, i.~e.~to find the ball of minimum radius which includes each of a set of balls in fixed dimension, Ref.~\cite{megiddoballs} presents an algorithm whose running time is linear in the number of balls. For the simpler case of $m$ equiprobable states represented by points, the problem of determining the Lagrange ball is the problem of finding the minimal enclosing ball, i.~e.~the ball with minimum radius which contains all points. Finding this minimal enclosing ball in $\mathbf{R}^3$, also known as the 1-center problem in $\mathbf{R}^3$, is an old \footnote{In 1857 the mathematician Sylvester posed this problem for a set of points in the plane.} and well-known problem in operations research. It was shown to be solvable with a linear time $O(m)$ algorithm in Ref.~\cite{megiddo}. 

For a general quantum state discrimination problem we can distinguish several types of states depending on their relation with the Lagrange operator $\sigma_{\mathcal{R}}$. We consider Eq. (\ref{useful}) in Theorem \ref{qsdd} which says that for all $x \in \mathcal{X}$, the POVM elements of an optimal measurement strategy are such that $E^x (\sigma_{\mathcal{R}}-\rho_x)=0$. Since $E^x \geq 0$ and $\sigma_{\mathcal{R}}-\rho_x \geq 0$, it implies that $E^x \leq P^x$ where $P^x$ is defined as the projector ($P^x={P^x}^2$) onto the kernel of $\sigma_{\mathcal{R}}-\rho_x$. 

\begin{figure}[htb]
\centering
\begin{tikzpicture}

\def \Xs {1cm}

\coordinate(O) at (0,0);\coordinate(A) at ($(O)+(90:\Xs)$);
\coordinate(B) at ($(O)+(0:\Xs)$); \coordinate(C) at
($(O)+(-160:\Xs)$); \coordinate(OO) at ($(O)+(0.1*\Xs,0.1*\Xs) $);

\node[below=1.1*\Xs,anchor=north] at (O){$(a)$};
\node[left,anchor=south east,blue] at (O){$S$};

{[blue,shorten >=1pt,shorten <=1pt,->] \node at (O)[draw,circle
through=(A)]{}; }

{[red, every node/.style={}] \node at (A)[above]{$A$}; \node at
(B)[right]{$B$}; \node at (C)[left]{$C$}; }

{[black!50]
\draw (A)--(B)--(C)--cycle; }

{[blue,shorten >=1pt,shorten <=1pt,->] \node at (O)[draw,circle
through=(A)]{}; \draw (O) to node[above,xshift=-3pt]{} (A); \draw
(O) to node[below]{} (B); \draw (O) to node[above,yshift=3pt]{} (C);
}

{[white] \foreach \point in {O,A,C,B} \fill (\point) circle (1pt); }
{[black!50] \foreach \point in {} \fill(\point) circle (0.5pt); }
{[red] \foreach \point in {A,C,B} \fill (\point) circle (0.5pt); }
{[blue] \foreach \point in {O} \fill (\point) circle (0.5pt); }

\coordinate(O) at ($(0,0)+(2.6*\Xs,0)$);\coordinate(A) at
($(O)+(45:\Xs)$); \coordinate(B) at ($(O)+(0:\Xs)$); \coordinate(C)
at ($(O)+(45+180:\Xs)$); \coordinate(OO) at
($(O)+(0.0*\Xs,0.15*\Xs)$);

\node[below=1.1*\Xs,anchor=north] at (O){$(b)$};
\node[left,anchor=south east,blue] at (O){$S$};

{[red,shorten >=1pt,shorten <=1pt,->] \node at (O)[draw,circle
through=(A)]{}; }

{[red, every node/.style={}] \node at
(A)[right,yshift=3pt,xshift=-2pt]{$A$}; \node at (B)[right]{$B$};
\node at (C)[left,yshift=-3pt,xshift=2pt]{$C$}; }

{[black!50]
\draw (A)--(B)--(C); }

{[blue,shorten >=1pt,shorten <=1pt,->] \node at (O)[draw,circle
through=(A)]{}; \draw (O) to node[above,xshift=-3pt]{} (A); \draw
(O) to node[below]{} (B); \draw (O) to node[above,yshift=3pt]{} (C);
}

{[white] \foreach \point in {O,A,C,B} \fill (\point) circle (1pt); }
{[black!50] \foreach \point in {} \fill(\point) circle (0.5pt); }
{[red] \foreach \point in {A,C,B} \fill (\point) circle (0.5pt); }
{[blue] \foreach \point in {O} \fill (\point) circle (0.5pt); }

\coordinate(O) at ($(0,0)+(5.2*\Xs,0)$);\coordinate(A) at
($(O)+(45:\Xs)$); \coordinate(B) at ($(O)+(0:0.7*\Xs)$);
\coordinate(C) at ($(O)+(45+180:\Xs)$); \coordinate(OO) at
($(O)+(0.0*\Xs,0.15*\Xs)$);

\node[below=1.1*\Xs,anchor=north] at (O){$(c)$};
\node[left,anchor=south east,blue] at (O){$S$};

{[blue,shorten >=1pt,shorten <=1pt,->] \node at (O)[draw,circle
through=(A)]{}; }

{[red, every node/.style={}] \node at
(A)[right,yshift=3pt,xshift=-2pt]{$A$}; \node at (B)[below]{$B$};
\node at (C)[left,yshift=-3pt,xshift=2pt]{$C$}; }

{[black!50]
\draw (A)--(B)--(C); }

{[blue,shorten >=1pt,shorten <=1pt,->] \node at (O)[draw,circle
through=(A)]{}; \draw (O) to node[above,xshift=-3pt]{} (A);
\draw (O) to node[above,yshift=3pt]{} (C); }

{[white] \foreach \point in {O,A,C,B} \fill (\point) circle (1pt); }
{[black!50] \foreach \point in {} \fill(\point) circle (0.5pt); }
{[red] \foreach \point in {A,C,B} \fill (\point) circle (0.5pt); }
{[blue] \foreach \point in {O} \fill (\point) circle (0.5pt); }

\end{tikzpicture}
\caption{S is the center of blue ball $\mathcal{B}_{\sigma}$ (here projected on the plane defined by the points $A$, $B$ and $C$) corresponding to the Lagrange operator $\sigma$. In (a) the triangle $ABC$ is acute-angled, and all three states $A$, $B$, $C$ are guessable. In (b) the triangle is right-angled and $B$ is nearly-guessable. In (c) the triangle is obtuse-angled. The interior ball defined by $A$ and $C$ includes the point $B$ and $B$ is unguessable.}
\label{triangle_fig}
\end{figure}
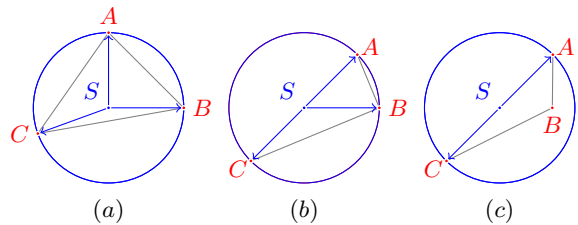

Let us first separate off a trivial case when there exists a $x_0 \in \mathcal{X}$ such that $\sigma_{\mathcal{R}}=\rho_{x_0}$ or ${\rm Ker}(\sigma_{\mathcal{R}}-\rho_x)=I_{\mathcal H}$. In this case, not measuring, but simply guessing the state to be $\rho_{x_0}$ is optimal (see also \cite{Hunter2}). Note that this case corresponds in the language of qubit cones to the case when the forward light cones of $\rho_x$ for $x \neq x_0$ all first intersect at $\rho_{x_0}$. If we are not in this trivial case, it is clear that for all $x$, $\rank(E^x) \leq \rank(P^x)\leq n-1$. Hence, this implies that for qubits, any optimal measurement strategy ${\cal E}=(E^x)$ has pure rank 1 POVM elements $E^x$. Now we can distinguish several types of states $\rho_x$:

There can be states $\rho_x$ for which $\dim {\rm Ker}(\sigma_{\mathcal{R}}-\rho_x)=0$, i.~e.~$E^x=P^x=0$. We will call such states {\em unguessable}, because no optimal strategy has $x$ as a possible outcome. For qubits, the geometric interpretation is that for such states $\rho_x$, the Lagrange ball $\mathcal{B}_{\sigma}$ is not tangent to the ball $\mathcal{B}_x$, i.e. the intersection $\mathcal{B}_{\sigma} \cap \mathcal{B}_x$ contains more than one point. An example is the point $B$ in Fig \ref{triangle_fig}(c). When $\rho_x$ is unguessable, we may remove $\rho_x$ from the set $\mathcal{R}$ without changing the Lagrange operator or the optimal strategies. 

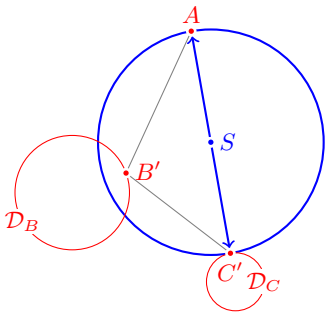
\begin{figure}[htb]
\centering
\begin{tikzpicture}

\def \Xs {15mm}

\coordinate(O) at (0,0);\coordinate(A) at ($(O)+(100:\Xs)$); \coordinate(B) at ($(O)+(180+100:\Xs)$); \coordinate(C) at ($(O)+(-160:0.8*\Xs)$); \coordinate(OO) at ($(O)+(-80:0.7*\Xs) $);
\node[right,blue] at (O){$S$};

{[blue,thick,shorten >=2pt,shorten <=2pt,->]
\node at (O)[draw,circle through=(A)]{};
\draw (O) to node[ left]{} (A);
\draw (O) to node[left]{} (B);
}

{[red, every node/.style={}] \node(nB) at (B)[circle, minimum
size=0.5*\Xs]{}; \node(nC) at (C)[circle, minimum size=1*\Xs]{};

\coordinate(BB) at (intersection  of nB and O--B); \node(Bb) at (BB)
[draw,circle through=(B)]{}; \coordinate(CC) at (intersection  of nC
and O--C); \node(Cc) at (CC) [draw,circle through=(C),red]{};

\coordinate(BBB) at (intersection of Bb and OO--BB);
\coordinate(CCC) at (intersection of Cc and OO--CC);

\node at (A)[above,yshift=0pt,xshift=0pt]{$A$};
\node at (Bb.0)[fill=white,inner sep=1pt]{$\mathcal{D}_C$};
\node at (Cc.-150)[color=red,fill=white,inner sep=1pt]{$\mathcal{D}_B$};
\node at (B) [below] {$C'$};
\node at (C) [right,yshift=1pt] {$B'$};
}

{[black!50]
\node at (OO)[below left]{};
\draw (B)--(C)--(A);
}

{[white]
\foreach \point in {O,A,B,C} \fill (\point) circle (2pt);
}
{[black!50]
\foreach \point in {} \fill(\point) circle (1pt);
}
{[red]
\foreach \point in {A,B,C} \fill (\point) circle (1pt);
}
{[blue]
\foreach \point in {O} \fill (\point) circle (1pt);
}

\end{tikzpicture}
\caption{The interior disk of $A$ and $\mathcal{D}_C$ intersects the disk $\mathcal{D}_B$. The vectors pointing from $S$ to the points $A$ and $C'$ represent the pure state vectors $E^x$ of the optimal strategy. The state $\rho_B$ is unguessable; its disk intersects the interior disk of $A$ and $\mathcal{D}_C$ and hence $E^B=0$. Note that the problem of distinguishing $\rho_A,\rho_B,\rho_C$ is equivalent to distinguishing the equiprobable states represented by the points $A$,$B'$ and $C'$. Thus the problem of distinguishing non-equiprobable states reduces to the problem of distinguishing equiprobable states in Fig.\ref{triangle_fig}(c).}
\label{obtusedisks}
\end{figure}

\begin{figure}[htb]
\centering
\begin{tikzpicture}

\def \Xs {15mm}

\coordinate(O) at (0,0); \coordinate(A) at ($(O)+(90:\Xs)$); \coordinate(B) at ($(O)+(-60:\Xs)$); \coordinate(C) at ($(O)+(-160:\Xs)$); \coordinate(OO) at ($(O)+(-80:0.75*\Xs) +(135:0.2\Xs)$);

\node[below left,xshift=3pt,blue] at (O){$S$};

{[blue,thick,shorten >=2pt,shorten <=2pt,->]
\node at (O)[draw,circle through=(A)]{};
\draw (O) to node[ left]{} (A);
\draw (O) to node[left]{} (B);
\draw (O) to node[below]{} (C);
}

{[red, every node/.style={}]
\node(nB) at (B)[circle, minimum size=0.5*\Xs]{};
\node(nC) at (C)[circle, minimum size=1*\Xs]{};
\coordinate(BB) at (intersection  of nB and O--B); \node(Bb) at (BB) [draw,circle through=(B)]{};
\coordinate(CC) at (intersection  of nC and O--C); \node(Cc) at (CC) [draw,circle through=(C),red]{};

\coordinate(BBB) at (intersection of Bb and OO--BB);
\coordinate(CCC) at (intersection of Cc and OO--CC);

\node at (A)[above,yshift=0pt,xshift=0pt]{$A$};
\node at (Bb.0)[fill=white,inner sep=1pt]{$\mathcal{D}_C$};
\node at (Cc.-150)[color=red,fill=white,inner sep=1pt]{$\mathcal{D}_B$};
\node at (B) [below] {$C'$};
\node at (C) [left] {$B'$};
}

{[black!50]
\node at (OO)[below left]{};
\draw (B)--(C)--(A)--cycle;
}

{[white]
\foreach \point in {O,A,B,C} \fill (\point) circle (2pt);
}
{[black!50]
\foreach \point in {} \fill(\point) circle (1pt);
}
{[red]
\foreach \point in {A,B,C} \fill (\point) circle (1pt);
}
{[blue]
\foreach \point in {O} \fill (\point) circle (1pt);
}

\end{tikzpicture}
\caption{In this example, the interior disk centered at $S$ is tangent to all disks and all states $\rho_A$, $\rho_B$ and $\rho_C$ are guessable. The vectors pointing from $S$ to the points $A$, $B'$ and $C'$ represent the pure state vectors $E^x$ of the optimal strategy.}
\label{acutedisks}
\end{figure}
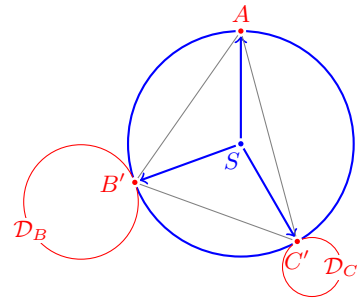

There can be states for which $1 \leq \dim {\rm Ker}(\sigma_{\mathcal{R}}-\rho_x) \leq n-1$. For qubits, this implies that the operator $E^x$ is proportional to the pure state projector onto this kernel, namely $E^x \propto P^x =I_{\mathcal{H}}+\frac{\vec{r}_x-\vec{s}}{t} \cdot \sigma$ with the Lagrange operator $\sigma\equiv (t,\vec{s})$. Geometrically, these projectors $P^x$ are vectors from the center $S$ of the Lagrange ball $\mathcal{B}_{\sigma}$ towards the center of the ball $\mathcal{B}_x$, see e.~g.~Figs.~\ref{triangle_fig}, \ref{obtusedisks}, \ref{acutedisks}. 
In order to determine an optimal strategy, one has to find nonnegative coefficients $\lambda_x$ such that for $E^x=\lambda_x P^x$, $\sum_x E^x=I_{\mathcal{H}}$. For qubits, the POVM completeness condition implies that $\sum_x \lambda_x \vec{r}_x=\vec{s}$ for weights $\lambda_x \geq 0$, $\sum_x \lambda_x=1$. It can be the case that for {\em all} optimal strategies $\lambda_x=0$, hence $E^x=0$. In this case we call $\rho_x$ nearly-guessable; an example is point $B$ in Fig. \ref{triangle_fig}(b). It is clear that when $\rho_x$ is nearly-guessable, it can be removed from the set of states without changing the Lagrange operator or the optimal strategies. In the last case when $E^x \neq 0$ for {\em some} optimal strategies $\mathcal{E}$, we will call $\rho_x$ guessable. \\


Now we present the procedure to discriminate between four states. It relies on checking whether the interior ball of increasingly larger subsets (from size 1 to 4) intersects with the remaining states. If this happens with a subset of size $k\leq 4$, then there is an optimal measurement strategy with these $k$ states as outputs or equivalently the $k$ states of the subset are all guessable. W.l.o.g. let $A$ be a point and $\mathcal{B}_{B,C,D}$ be three balls. 

{\bf 1}. The point $A\in\mathcal{B}_B\bigcap\mathcal{B}_C\bigcap\mathcal{B}_D$ in which case the interior ball is point $A$; this is the trivial case where just guessing $A$ is one optimal strategy. 



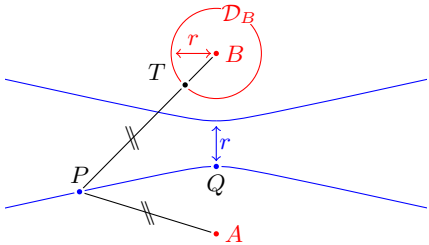
\begin{figure}[htb]
\centering
\begin{tikzpicture}[every node/.style={font=\small}]
\def \Xs {2cm}

\coordinate(O) at(0,0); \coordinate (B) at (0,0.7*\Xs); \coordinate(A) at (0,-0.5*\Xs); 
\node(BB) at (B)[circle, draw=red, minimum size=0.6*\Xs] {}; \coordinate(O0) at ($0.5*(A)+0.5*(B)$); \node[red,inner sep=0.2pt,fill=white] at (BB.60){$\mathcal{D}_B$};
\coordinate(P0) at ($0.5*(A)+0.5*(BB.south)$); \coordinate(P1) at ($0.5*(A)+0.5*(BB.north)$);

\coordinate(T1) at (tangent cs:node=BB,point={(A)},solution=2); \coordinate(O1) at ($0.5*(T1)+0.5*(A)$); \coordinate(NW) at ($(O0)+20*(O1)-20*(O0)$); 
\coordinate(T2) at (tangent cs:node=BB,point={(A)},solution=1); \coordinate(O2) at ($0.5*(T2)+0.5*(A)$); \coordinate(SW) at ($(O0)+20*(O2)-20*(O0)$);

\coordinate(SE) at ($(O0)-20*(O1)+20*(O0)$);\coordinate(NE) at ($(O0)-20*(O2)+20*(O0)$);

\clip ($(O)+(-1.4*\Xs,-0.6*\Xs)$) rectangle ($(O)+(1.4*\Xs,1*\Xs)$);

{[black!50]
\draw[blue] (P1) .. controls ($(P1)+(180:0.3)$)  .. ($(NE)$);
\draw[blue] (P1) .. controls ($(P1)+(0:0.3)$)  .. ($(SE)$); 
}

{[black]
\draw[blue] (P0) .. controls ($(P0)+(180:0.3)$) .. ($(NW)$) coordinate[pos=0.65](P); \coordinate(T) at (intersection 2 of BB and P--B);
\draw[blue] (P0) .. controls ($(P0)+(0:0.3)$)  .. ($(SW)$);
\draw (A)--node{$\backslash\!\!\backslash$}(P)--node{$\backslash\!\!\backslash$}(T);\draw[<->, shorten <=2pt,shorten >=2pt,red] (BB.west)--node[above]{$r$}(B);
\draw (T)--(B);
\draw[<->, shorten <=2pt,shorten >=2pt,blue] (P0)--node[right,inner sep=1pt]{$r$}(P1);
}

\node[right,red] at (A) {$A$};\node[right,red] at (B) {$B$};\node[above] at (P) {$P$};\node[left,xshift=-4pt,yshift=5pt] at (T) {$T$};\node[below] at (P0) {$Q$};

\foreach \point in {A,B,P,T,P0}
	\fill[white] (\point) circle (2pt);
\foreach \point in {A,B}
	\fill[red] (\point) circle (1pt);
\foreach \point in {P,P0}
	\fill[blue] (\point) circle (1pt);
\foreach \point in {T}
	\fill[black] (\point) circle (1pt);
		
\end{tikzpicture}
\caption{Let $r$ be the radius of the disk $\mathcal{D}_B$. The point $Q$ is the closest point to $A$ which is equidistant to the red circle defined by the disk $\mathcal{D}_B$. All points equidistant from $A$ and the circle form a hyperbola whose foci are $A$ and $B$, i.e. $|\overline{AP}|=|\overline{PT}|$. In general, the points on a hyperbola are such that the difference between the distance to the foci $A$ and $B$ is constant. In this case the constant is $r$.}
\label{hyperbole}
\end{figure}

{\bf 2}. The interior ball of a pair of balls, provided that it is unique, has a non-empty intersection with the third {\em and} the fourth ball. Then, this interior ball is the interior ball of the whole set. Clearly, it is sufficient to check for pairs of balls such that their interior ball is unique, since the Lagrange operator ball itself is unique. Note that it is very easy to compute the interior ball of a pair of balls by placing the center $S$ of the interior ball midway between the two balls. An example of this scenario is depicted in Fig. \ref{obtusedisks} when distinguishing three states. Note that in this case, it can happen that the third and fourth state are also guessable, for example any two antipodal points which are on the blue circle in Fig.~\ref{obtusedisks} would also be guessable. If this is not the case, then we consider whether:


{\bf 3}. The interior ball of a triplet of balls, provided that it is unique, has a non-empty intersection with the fourth ball. Then, again this interior ball of this subset is the interior ball of the whole set. We can compute the interior ball of a triplet of balls, say, $\mathcal{B}_A$, $\mathcal{B}_B$ and $\mathcal{B}_C$ as follows. Note that all three balls may have a radius strictly larger than zero; if this is so, then we redefine the balls as $\mathcal{B'}_{A,B,C}$ with radius 
$|p-p_{x=A,B,C}|$ where $p=\max_{x=A,B,C} p_x$ in order to simplify the description of the following procedure. It is clear that we can restrict ourselves to the plane defined by the centers $A$, $B$ and $C$ and hence we represent the balls by disks $\mathcal{D}_{x=A,B,C}$ where w.l.o.g we assume that $\mathcal{D}_A$ is a point. The interior ball $\mathcal{B}_{\sigma}$ can be represented as the interior disk $\mathcal{D}_{\sigma}$ (i.~e.~the disk with minimum radius which intersects all disks $\mathcal{D}_x$). 
Now, we can first check whether or not the interior disk of a subset of two states intersects the third state; we already did this computation in Case {\bf 2} and use the results here. If there is no subset of two states whose interior disk includes the third one, then all three states are guessable and we determine the interior disk as follows:

We find the center $S$ of the Lagrange operator disk which is the closest point which is equidistant from the disks and the point. Thus, first we find the set of points which are equidistant from the point $A$ and the surface of the disk $\mathcal{D}_B$, this set of points form a hyperbola $\mathcal{H}_{AB}$ described by a quadratic equation, see Fig. \ref{hyperbole}. Similarly, we find the points which are equidistant from $A$ and the disk $\mathcal{D}_C$, the hyperbola $\mathcal{H}_{AC}$. Computing the intersection $\mathcal{H}_{AB}\bigcap\mathcal{H}_{AC}$ requires solving a cubic equation and the intersection will consists of at most four points. The interior disk with smallest radius will have its center $S$ at the point closest to $A$ and its radius is $|\overline{AS}|$, see the example of Figure \ref{acutedisks}.

Once we have found the Lagrange operator or the Lagrange ball of a subset of three states, we can check whether it includes the fourth state. If no three-state subset defines the Lagrange operator, we have as the last possibility:

{\bf 4}. The interior ball of the states is tangent to the four balls $\mathcal{B}_A$, $\mathcal{B}_B$, $\mathcal{B}_C$ and $\mathcal{B}_D$ and we determine the center and radius of this ball as follows. The points which are equidistant from $A$ and the surface of $\mathcal{B}_{x=B}$ (resp. $\mathcal{B}_{x=C}$ and $\mathcal{B}_{x=D}$) form a hyperboloid of revolution (of two sheets) $\mathcal{H}_{AB}$ (resp. $\mathcal{H}_{AC}$ and $\mathcal{H}_{AD}$) in three dimensions with the foci $A$ and $x=B$ (resp. $x=C$ and $x=D$). The line-segment of the major axis of $\mathcal{H}_{Ax}$ from one hyperboloid sheet to the other is of length $|\vec{r}_x|$ for $x=B,C,D$ (see the line-segment of length $r$ in Fig.~\ref{hyperbole}). 

We can compute the intersection of these three (quadratic) hyperboloids $\mathcal{H}_{AB}$,$\mathcal{H}_{AC}$,$\mathcal{H}_{AD}$ which consists of at most 8 points. The computation of the intersection is a task which can be reduced to solving two quartic equations. Again we choose the point of intersection which is closest to $A$; this is the center $S$ of the Lagrange ball and its radius is $|\overline{AS}|$.\\


{\em Remark}: Note that for equiprobable states which can be represented as points, the boundary of the minimal enclosing ball in Case 2, 3 and 4 is the circumscribed sphere of respectively 2, 3 and 4 points.

In conclusion, we have analyzed the problem of qubit state discrimination and have given an analytical procedure for explicitly solving the problem of distinguishing between four arbitrary states. This procedure can be easily checked to give the answers derived in Ref.~\cite{mirror} for the problem of distinguishing three mirror-symmetric states (see \cite{deconinck}). It is clear that the linear-time algorithm and the procedure to compute the Lagrange operator for a set of four qubit states are particular to qubits and do not directly generalize to higher dimensions where the constraints of positive semi-definiteness are no longer quadratic. However, an interesting question for future research is to analyze whether the SDP for the Lagrange operator, Eq.~(\ref{qsdd}), is sufficiently simple in its dependence on $\sigma$ and $\rho_x$ to allow for some of the search techniques in \cite{megiddoballs,megiddo} to apply. In addition, the notion of guessable, nearly-guessable and unguessable states and the observation about the number of outcomes for an optimal strategy are general and can also be useful in describing the optimal strategies for general quantum state discrimination problems.  

BMT acknowledges support by the DARPA QUEST program under contract number HR0011-09-C-0047. MED acknowledges support by the department INFRES of TELECOM ParisTech.

\end{document}